\documentclass[aps,prd,onecolumn,groupedaddress,showpacs,nofootinbib,amssymb
]{revtex4}
\usepackage[dvips]{graphicx}
\usepackage{amssymb}
\usepackage{amsmath}
\usepackage{graphicx,,color}
\usepackage{amsfonts}
\usepackage{bm}
\usepackage{cancel}
\usepackage{comment}

\allowdisplaybreaks[4]

\begin{document}

\tolerance=5000

\title{Constraints on Cosmic Opacity from Bayesian Machine Learning: \\ The hidden side of the $H_{0}$ tension problem}

\author{
Emilio Elizalde$^{1,2}$\footnote{E-mail: elizalde@ieec.uab.es} and
Martiros Khurshudyan$^{1,2}$\footnote{Email: khurshudyan@ice.csic.es}}
\affiliation{
$^{1}$Consejo Superior de Investigaciones Cient\'{\i}ficas, ICE/CSIC-IEEC,
Campus UAB, Carrer de Can Magrans s/n, 08193 Bellaterra (Barcelona) Spain \\
$^{2}$International Laboratory for Theoretical Cosmology, Tomsk State University of Control Systems and Radioelectronics (TUSUR), 634050 Tomsk, Russia 
}

\begin{abstract}

Bayesian (Probabilistic) Machine Learning is used to probe the opacity of the Universe. It relies on a generative process where the model is the key object to generate the data involving the unknown parameters of the model, our prior beliefs, and allows us to get the posterior results. The constraints on the cosmic opacity are determined for two flat models, $\Lambda$CDM and XCDM (this having $\omega_{de} \neq -1$), for three redshift ranges, $z\in[0,2.5]$, $z \in[0,5]$, and $z\in[0,10]$, in each case. This is to understand how the constraints on the cosmic opacity could change in the very deep Universe, and also to check to what extent there is a redshift-range dependence. The  following forms for the opacity, $\tau(z) = 2\epsilon z$ and $\tau(z) = (1+z)^{2\epsilon} -1$, corresponding to an observer at $z=0$ and a source at $z$, are considered. The results of our analysis show that the Universe is not fully transparent, and this may have a significant impact on the $H_{0}$ tension problem. In the generative process, the fact that, owing to cosmic opacity, the flux received by the observer is reduced has been taken into account. In the analysis, the luminosity distance associated with the cosmological model has been employed.
\end{abstract}

\pacs {}

\maketitle

\section{Introduction}\label{sec:INT}

Machine Learning algorithms have proven to be very efficient for analyzing a large amount of data and, specifically, for finding patterns in them, significantly speeding up the study of various different processes in physics, chemistry, medicine, and biology. Recently, Machine Learning has been used in cosmology and astrophysics, showing clear advantages over other more usual methods employed in similar studies. In this regard, it is natural to explore the limits that Machine Learning can face, since a huge amount of observational data will be accumulated in the near future, and an efficient  data analysis at such huge scales will become a key aspect of the research in cosmology and astrophysics. Therefore it is necessary to start testing these methods now, even if the amounts of data are, at present, not that large. Indeed, owing to its predictive power, Machine Learning can be already used to design new experiments or observational tests,  allowing also to follow  specific paths in a given observation or experiment. To understand how Machine Learning algorithms can be used, say in cosmology, let us explore in more detail what is behind these methods. In general, any Machine Learning algorithm consists of the following three steps. First, one has to define the model. Then, we need to choose a set of data and run the learning algorithm; this means that we choose the structure of the neural network and the way to do the optimization, to find the unknown weights charactering the network~\cite{MLStart}~-~\cite{MLEnd}~(to mention a few). On the other hand, to define the model, one usually defines a set of equations or rules describing the model behavior. Now, what about the data?  The data, in this case, will be the information obtained  from some experiment or observation. Further, by displaying the data in terms of input and output pairs, one runs some optimization algorithm, in order to get a final set of weights giving the specific neural network. In other words, we train the neural network and make it ready to perform some predictions and determine or constrain the free parameters of the model. 

However, there is another interesting approach that can be used instead of the usual Machine Learning procedures: it is known as the Bayesian Machine Learning or Probabilistic Machine Learning. The steps to follow are the same as in the case of Machine Learning, however the definition of the data has to be modified. The main goal of our study is to apply a Bayesian (Probabilistic) Learning in order to obtain the constraints on the cosmic opacity function. More details about the models and the steps behind Bayesian (Probabilistic) Learning will be given in Sect.~\ref{sec:Models}. It will be shown there, that our approach is original and different as in the case of former studies constraining cosmic opacity~\cite{OPStart}~-~\cite{OPEnd}. Now let us come back to the basic discussion on cosmic opacity and on why its consideration is important for cosmology. 

The whole issue starts from the crucial discovery of the accelerated expanion of the Universe, through the observed (and fully unexpected) dimming of type Ia supernovae~(SNe Ia)~\cite{Riess},~\cite{Perlmutter}. Trying to understand what is behind the mechanism accelerating the expansion brings about several interesting ideas. A key one is the existence of a dark energy, and the way how dark energy can be introduced into the field equations of General Relativity describing the background dynamics. One  option is to consider it as a fluid~\cite{Bamba:2012cp}~-~\cite{Arman}\footnote{Ref.-s~\cite{Bamba:2012cp}~-~\cite{Arman} provides a comprehensive discussion about fluid dark energy and related problems. They are useful for future discussions on alternative ways to present dark energy, including also modification of General Relativity.}. However, there are other phenomena which may contribute to explaining why type Ia supernovae (SNe Ia) are observed to be fainter than expected. Indeed, the SNe Ia observations are affected by at least four different possible sources of opacity. Namely,  the Milky Way dust, the intergalactic medium, the intervening galaxies or the host galaxy opacity. In this regard, it should be mentioned that the issue, to which extent is the Universe transparent, may have a significant effect on the model parameter estimates and lead to a different result\footnote{More details can be found in ~\cite{OPStart}~-~\cite{OPEnd}.}. On the other hand, we still need to estimate whether other regions of the Universe are transparent, or not, since subsequent observations independently have confirmed the accelerated expansion of the Universe~\cite{Ade1}~-~\cite{Hinshaw}. In the literature, we learn that the tests of cosmic opacity focus on the cosmic distance duality~(CDD) relation~\cite{OPStart}~-~\cite{OPEnd}
\begin{equation}
\frac{D_{L}}{D_{A}} (1+z)^{-2} = 1,
\end{equation}
where $D_{L}$ and $D_{A}$ are the luminosity distance and the angular diameter distance, respectively. On the other hand, it is well known that this relation is valid for all cosmological models based on Riemannian geometry and that the number of photons is conserved. Therefore, first of all, the reason for the violation of the CDD relation is the non-metric theory of gravity and the fact that the photons do not follow ordinary null geodesics. On the other hand, it could hint towards non-conservation of the number of photons. However, under the assumption that photon traveling along null geodesics is a more fundamental and unassailable one, then we are left only with the option that the violation of the CDD relation is most likely due to non-conservation of the photon number, which can be related to the presence of some opacity source and, also, to some non-standard exotic physics. In this case, the flux detected by the observer will be reduced, and the observed luminosity distance will be
\begin{equation}\label{eq:RDDR}
D_{L,obs} = D_{L,true} e^{\tau/2},
\end{equation}
where $\tau$ is an opacity parameter denoting the optical depth associated with the cosmic absorption. In the recent literature, there are many papers investigating cosmic opacity, assuming that the non-conservation of the photon number is the reason for the violation of the relation. From the observational point of view, if the Universe is opaque, the luminosity distance measurements will be affected significantly. Therefore, the best way to test the cosmic opacity observationally is the independent measurements of the intrinsic luminosities and sizes of the same object, without using any specific cosmological model. However, we should take into account that, due to astrophysical complications and instrumental limitations, observations of this type are extremely challenging. 

On the other hand, standard sirens can play an important role helping to develop alternative tools for indicating opacity-free distances. It is due to the fact that in FLRW metric, Gravitational Waves propagate freely through a perfect fluid without any absorption and dissipation. Therefore, the confrontation of the luminosity distance derived from SNe Ia with  measures from Gravitational Waves sources, can yield a new scheme to investigate the  Universe opacity. However, the redshift mismatch between Gravitational Waves events and the SNe Ia sample still remains a challenge, too. What are the ways to overcome all these observational difficulties, and what are the limits on the cosmic opacity studies? The answer to these questions will depend on the quality of the observational data, which should be increased. However, for the moment we can already overcome some of the difficulties, by using Machine Learning algorithms. 

In this paper we will prove that the cosmic opacity can be studied and proper constraints on the free parameters of the model  can indeed be obtained provided we adopt the Bayesian (Probabilistic) Learning approach, where the data are generated from Eq.~(\ref{eq:RDDR})\footnote{Actually, in the analysis we use $D^{2}_{L,obs} = D^{2}_{L,true} e^{\tau}$.}. It should be mentioned that, in Eq.~(\ref{eq:RDDR}), $D_{L,true}$ is associated with the specific cosmological model under consideration. In other words, using the Bayesian Learning approach we will constrain, not only the cosmic opacity, but  also the parameters of the model used to calculate $D_{L,true}$. The method and the models will be discussed in Sect.~\ref{sec:Models}, where it will be seen that we can obtain constraints, making our study original and unique, as compared to other studies reported in the recent literature. On the other hand, in order to compare our results with previous studies on the same topic we consider the following two particular parameterizations $\tau(z) = 2\epsilon z$ and $\tau(z) = (1+z)^{2\epsilon} -1$, which describe the phenomenological forms of  $\tau(z)$ corresponding to the opacity between an observer at $z=0$ and a source at $z$. The detailed motivation for considering these two forms for $\tau(z)$, and for using the relation given by Eq.~(\ref{eq:RDDR}) for the generative process, can be found  in~\cite{OPStart}~-~\cite{OPEnd}.

The paper is organized as follows. Details on the method used in the paper and on the cosmological model description are given in Sect.~\ref{sec:Models}. Sect.~\ref{sec:NRAD} deals with the constraints on the two different parameterizations of the cosmic opacity, obtained from Bayesian (Probabilistic) Learning for each of the phenomenological $\tau(z)$ forms considered, corresponding to the opacity between an observer at $z=0$ and a source at $z$. Our results show that the Universe is not fully transparent. The constraints on the free parameters describing the $\Lambda$CDM and XCDM models, obtained with the same procedure, are  presented in the second part of Sect.~\ref{sec:NRAD}. Finally,  Sect.~\ref{sec:conc} is devoted to conclusions.

\section{The method and the models}\label{sec:Models}

In this section we describe the philosophy behind the Bayesian (Probabilistic) Machine Learning method, a likelihood-free inference approach, shedding light on its crucial points, mainly. The precise implementation of the procedure is actually a personal issue, since different programming codes can be involved; therefore, we will not display a specific code to be used, or a way how the data can be generated, or how exactly the priors can be updated. Instead, we use the PyMC3 Python-based framework providing the necessary tools to perform the analysis in such a way that we could solely concentrate our attention on the problem. On the other hand, in the second part of this section we will discuss the cosmological models considered in order to calculate the luminosity distance employed in the data generation. Each discussion will be organized in a separate subsection, for clarity and better identification.

\subsection{Method}

As we have already mentioned, the goal of this work is to use Bayesian (Probabilistic) Machine Learning in order to determine the opacity affecting luminous signals coming from different domains of the Universe. In this section we  discuss  the physical background behind this method and what are the cosmological models preferably used to calculate the luminosity distance, which eventually lead to the constraints on the cosmic opacity. We want to keep our discussion as simple as possible since, in our opinion, the main issue here is to clearly understand the philosophy behind the approach; its implementation in terms of a specific code in a particular computation language may be done in different (equivalent) ways, depending on personal taste and abilities. This just means that someone will prefer to employ his/her personal code, while others may prefer to use one of the already available frameworks, concentrating attention on the problem itself. We take the second alternative and will use, in our study, the PyMC3~\cite{PyMC3}\footnote{https://docs.pymc.io} framework. This is a python-based framework, endowed with a comprehensive set of pre-defined statistical distributions that can be used to build the models. It should be mentioned that it makes use of Theano\footnote{http://deeplearning.net/software/theano} - a deep learning python-based library, to construct probability distributions to implement cutting edge inference algorithms. 

We have chosen to work with PyMC3, because it allows to write down models using an intuitive syntax to describe the data generating process. On the other hand, it allows to involve gradient-based MCMC algorithms, for fast approximate inference, or to use Gaussian processes in order to build Bayesian non-parametric models. Indeed, by using PyMC3 in our study, we have realized that it does provide all necessary tools for the analysis, allowing to concentrate attention  on the actual problem, only. Of course, there are other available frameworks, as well. In future studies we plan to compare the results from these alternative frameworks, to try to understand the limitations and advantages imposed by each of them. 

Having chosen the tools to be used in our analysis, we will now discuss some crucial aspects of Bayesian (Probabilistic) Machine Learning. As in other other Machine Learning algorithms, one has to proceed through the following steps:
\begin{itemize}
\item First, to define the model.  To this end we will use a so-called generative process. This means that the model is employed to generate the data: a sequence of steps describing how the data were created will be defined. On the other hand, since we have a generative process, we have to incorporate our prior beliefs to the unknown parameters describing the model. 

\item The second one is the most crucial step: to define our Bayesian (Probabilistic) Machine Learning algorithm. What makes it different from other Machine Learning methods is the data interpretation. While in the case of a usual Machine Learning algorithm, one would use the data from some astronomical observation or experiment, here the data to be used is obtained from the generative process.

\item Finally, the last step is to run the learning algorithm, update our belief about the parameters and get a new distribution over them. In other words, we define a systematic way to impose and improve the constraints on the  parameters of the model. 
\end{itemize}

From the above description, we observe that Bayesian (Probabilistic) Machine Learning using forward simulation provides constraints from the simulated data-sets. Therefore, there is no meed to evaluate the corresponding likelihood function. Bayesian (Probabilistic) Machine Learning can be interesting for data analysis problems where complex physical processes and instrumental effects can be advantageously simulated; but incorporating them into the likelihood function and solving the inverse inference problem is a much harder task. This opens the door to a new paradigm for simulation-based analysis in cosmology, astrophysics, and other research fields. In the next section we will see that it can be used to constrain cosmic opacity for a given cosmological model, by using the generative process based on Eq.~(\ref{eq:RDDR}).

\subsection{Models}

In concomitance with previous analysis of cosmic opacity, we will concentrate our attention on two cosmological models, namely the flat $\Lambda$CDM one and an XCDM model. It is  well known that a flat XCDM model yields the following expansion rate
\begin{equation}\label{eq:HXCDM}
H(z) = H_{0} E(z,\Omega_{dm},\omega_{de}),
\end{equation}
with 
\begin{equation}\label{eq:EXCDM}
E(z,\Omega_{dm},\omega_{de}) = \left [ \Omega_{dm} (1+z)^{3} + (1- \Omega_{dm}) (1+z)^{3(1+\omega_{de})}\right]^{1/2},
\end{equation}
where $H_{0}$ and $\Omega_{dm}$ are the Hubble parameter and the fraction of dark matter at $z=0$, while $\omega_{de}$ is the equation of state parameter describing dark energy, respectively. On the other hand, if we set $\omega_{de} = -1$,  the XCDM model reduces to usual $\Lambda$CDM, where dark energy is given by the cosmological constant. Now, as follows from the discussion in the subsection above, we need to define the model for the generative process allowing to generate the "observational" data to be used in the analysis. In our case, we use $D^{2}_{L,obs} = D^{2}_{L,true} e^{\tau}$, with two phenomenological forms for $\tau(z)$~(see for instance Ref.~\cite{OPStart}), namely
\begin{equation}\label{eq:tau1}
\tau(z) = 2 \epsilon z,
\end{equation}
and
\begin{equation}\label{eq:tau2}
\tau(z) = (1+z)^{2\epsilon} -1,
\end{equation}
where $\epsilon$ is the parameter to be fitted. As the last step to complete the generative process,  based on Eq.~(\ref{eq:RDDR}), we need to associate $D_{L,true}$ with the cosmological models used in this paper that would be calculated following the definition of the luminosity distance. In other words, we calculate it in the following way
\begin{equation}\label{eq:DLTrue}
D_{L, theory} = (1+z) c \int_{0}^{z} {\frac{dz^{\prime}}{H(z^{\prime})}},
\end{equation}  
where $H(z)$ is the expansion rate given by Eqs.~(\ref{eq:HXCDM}) and (\ref{eq:EXCDM}). In this regard, it should be mentioned that our procedure allows to set constraints, not only on the cosmic opacity but also on the parameters of the underlying cosmology. In other words, based on the generative process, which is the key element of this Bayesian (Probabilistic) Machine Learning approach, starting from Eq.~(\ref{eq:RDDR}) we will constrain the cosmological model and the associated cosmic opacity. That is, in total we will constrain $4$ parameters and $3$ parameters, for the XCDM and $\Lambda$CDM models, respectively. The following aspect, specific of our analysis, makes it quite powerful and unique among similar studies: we can obtain constraints on the cosmic opacity for any given redshift range, even in the absence of corresponding astronomical data, since the data used in our analysis are the outcome of the generative process. Taking into account this fact, we constrain the models for the three following redshift ranges, namely $z\in [0,2.5]$, $z\in [0,5]$, and $z\in [0,10]$. In the next section we will display and discuss the results obtained in our investigation.

\section{Numerical results}\label{sec:NRAD}

In this section,  the constraints obtained using our Bayesian (Probabilistic) Learning approach, as discussed in Sect.~\ref{sec:Models}, will be presented, under the form of several plots\footnote{The analysis is based on $10$ chains and in each chain, $5000$ "observational" data-sets from the models have been simulated/generated.}. In order to lighten the presentation, some of the plots have been moved to an Appendix,  with proper description headings. We start with the discussion of the case when the ordinary $\Lambda$CDM model is used in the generative process. 

\subsection{Cosmic opacity with $\Lambda$CDM}

\begin{figure}[h!]
 \begin{center}$
 \begin{array}{cccc}
\includegraphics[width=90 mm]{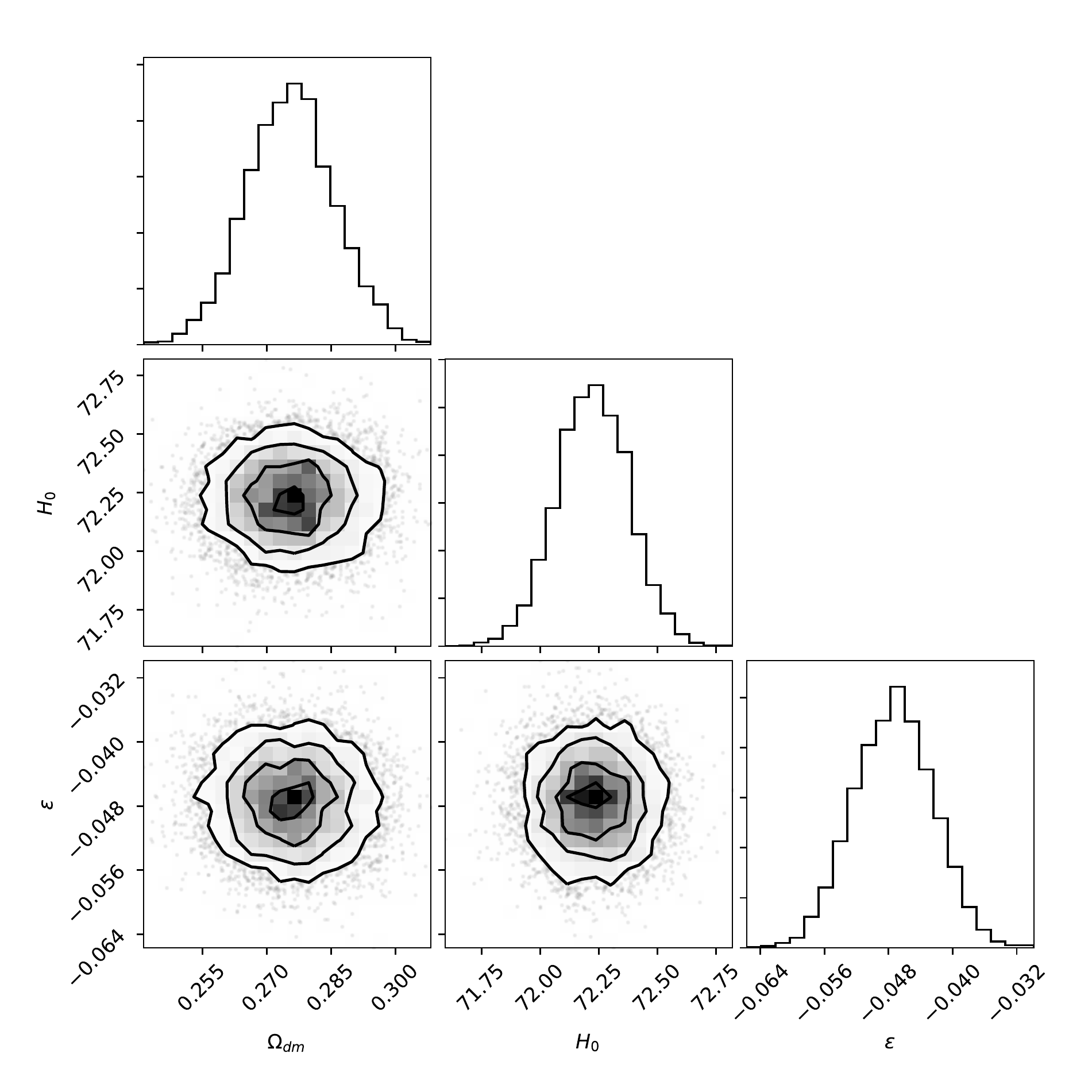}
\includegraphics[width=90 mm]{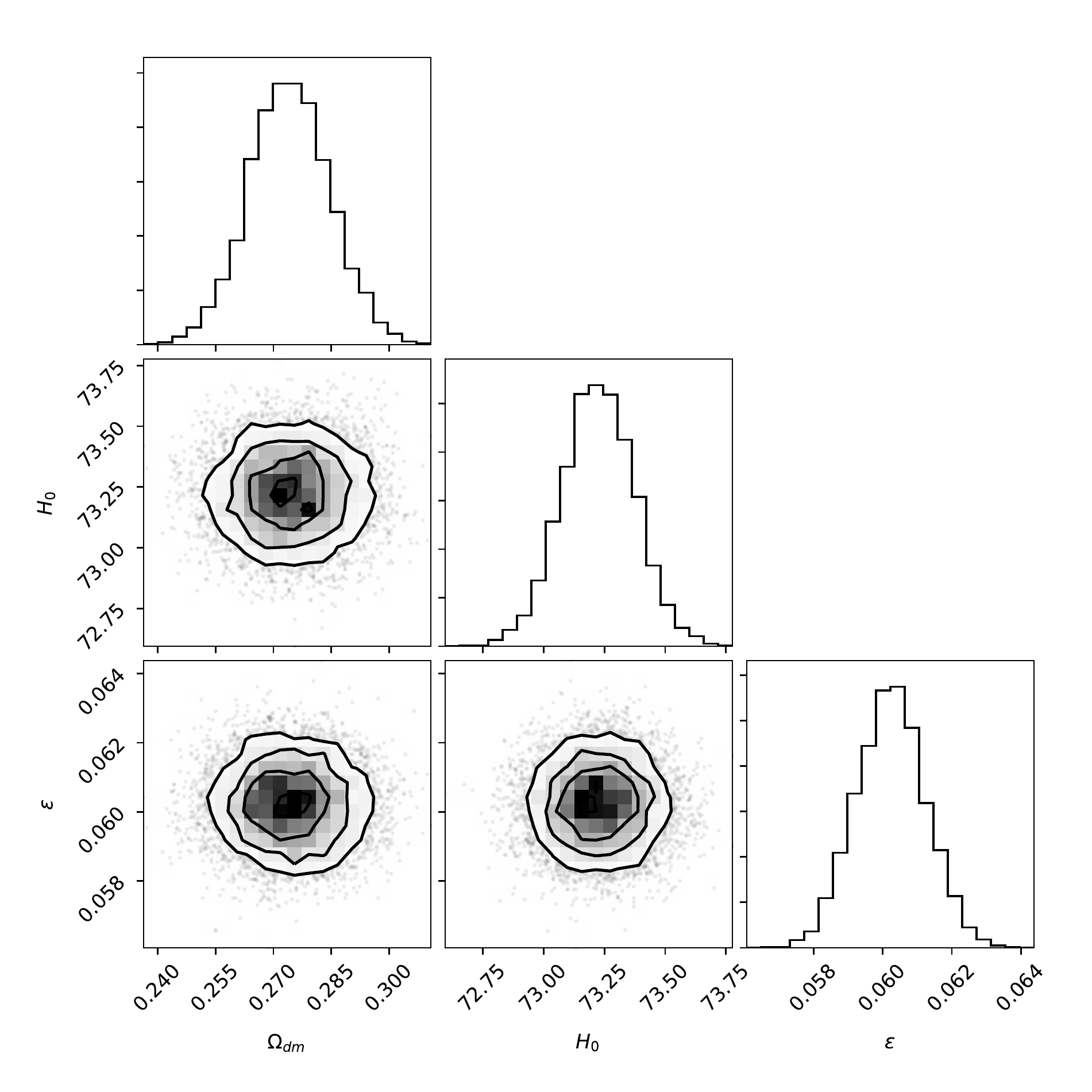}
 \end{array}$
 \end{center}
\caption{The contour map of the model given by Eq.~(\ref{eq:tau1}) for $z \in [0,2.5]$ is given in the left hand side plot. The right hand side plot corresponds to the contour map of the same model for $z \in [0,5]$. The best fit values of the model parameters, when $\tau(z)$ is given by  Eq.~(\ref{eq:tau2}), are $\Omega_{dm} = 0.276 \pm 0.01$, $H_{0} = 72.23 \pm 0.15$ and $\epsilon = -0.04 \pm 0.005$ for $z\in[0,2.5]$, while for $z \in [0,5]$ they are found to be $\Omega_{dm} = 0.274 \pm 0.01$, $H_{0} = 73.23 \pm 0.146$ and $\epsilon = 0.0602 \pm 0.001$. The theoretical calculations of the luminosity distance are for the $\Lambda$CDM model.}
 \label{fig:Fig1_1}
\end{figure}

The study of the cosmic opacity in the case of $\Lambda$CDM requires to constrain the three parameters $H_{0}$, $\Omega_{dm}$ and $\epsilon$. Starting from Eqs.~(\ref{eq:HXCDM}), (\ref{eq:EXCDM}), and (\ref{eq:DLTrue}) and taking $\omega_{de} = -1$, we run the generative process with PyMC3, using Eq.~(\ref{eq:RDDR}) with $\tau(z)$ from Eq.~(\ref{eq:tau1}). The results of our analysis can be summarized as follows:

\begin{itemize}

\item The constraints $\Omega_{dm} = 0.276 \pm 0.01$, $H_{0} = 72.23 \pm 0.15$ and $\epsilon = -0.04 \pm 0.005$ for $z\in[0,2.5]$ have been obtained. We choose this redshift range because it covers the  range of observed $H(z)$ data. It should be mentioned that the measurement of the expansion rate does not provide the measurements of the distances, however, it can be used to validate the results obtained from Bayesian (Probabilistic) Machine Learning method. It is, therefore, interesting to consider this redshift range. We observe that the Bayesian (Probabilistic) Machine Learning approach puts very tight constraints on $\Omega_{dm}$ and $\epsilon$. Also, the very tight constrain on $\epsilon$ shows that it is not possible to have a fully transparent Universe. The contour map corresponding to this case is depicted in Fig.~(\ref{fig:Fig1_1})~(left plot).

\item The  generative process for $z\in[0,5]$ gives the following constraints - $\Omega_{dm} = 0.274 \pm 0.01$, $H_{0} = 73.23 \pm 0.146$ and $\epsilon = 0.0602 \pm 0.001$. As we can see, we get again very tight constraints on the parameters and the Universe cannot be fully transparent, either. The contour map corresponding to this case is depicted in Fig.~(\ref{fig:Fig1_1})~(the right-hand side plot). We notice that  the mean value of $H_{0}$ coming from the analysis hints to a solution of the $H_{0}$ tension problem, showing  that a proper analysis of the opacity can shed light on this problem\footnote{The $H_{0}$ tension problem arose from the Planck and Hubble Space Telescope result reports indicating that there is a huge difference between the calculated values of $H_{0}$ obtained from the two observational data sets. In the recent literature, there is an intensive discussion on how the problem can be solved, see~\cite{H0Start}~-~\cite{H0End}}.

\item On the other hand, the Bayesian (Probabilistic) Machine Learning approach sets the constraints $\Omega_{dm} = 0.289 \pm 0.009$, $H_{0} = 73.26 \pm 0.144$ and $\epsilon = 0.099 \pm 0.001$,  when $z\in[0,10]$. We observe that the status of the non-transparent Universe remains the same as in the case of the previous redshift ranges, considered above. However, we obtain a slightly high mean value for $\Omega_{dm}$, and thus the $H_{0}$ tension is again alleviated.  The range  $z\in[0,10]$ has been considered since it may be easily covered by GRB samples. The contour map corresponding to this case is given in Fig.~(\ref{fig:Fig1_3})~(the left-hand side plot).

\end{itemize}

\begin{figure}[h!]
 \begin{center}$
 \begin{array}{cccc}
\includegraphics[width=90 mm]{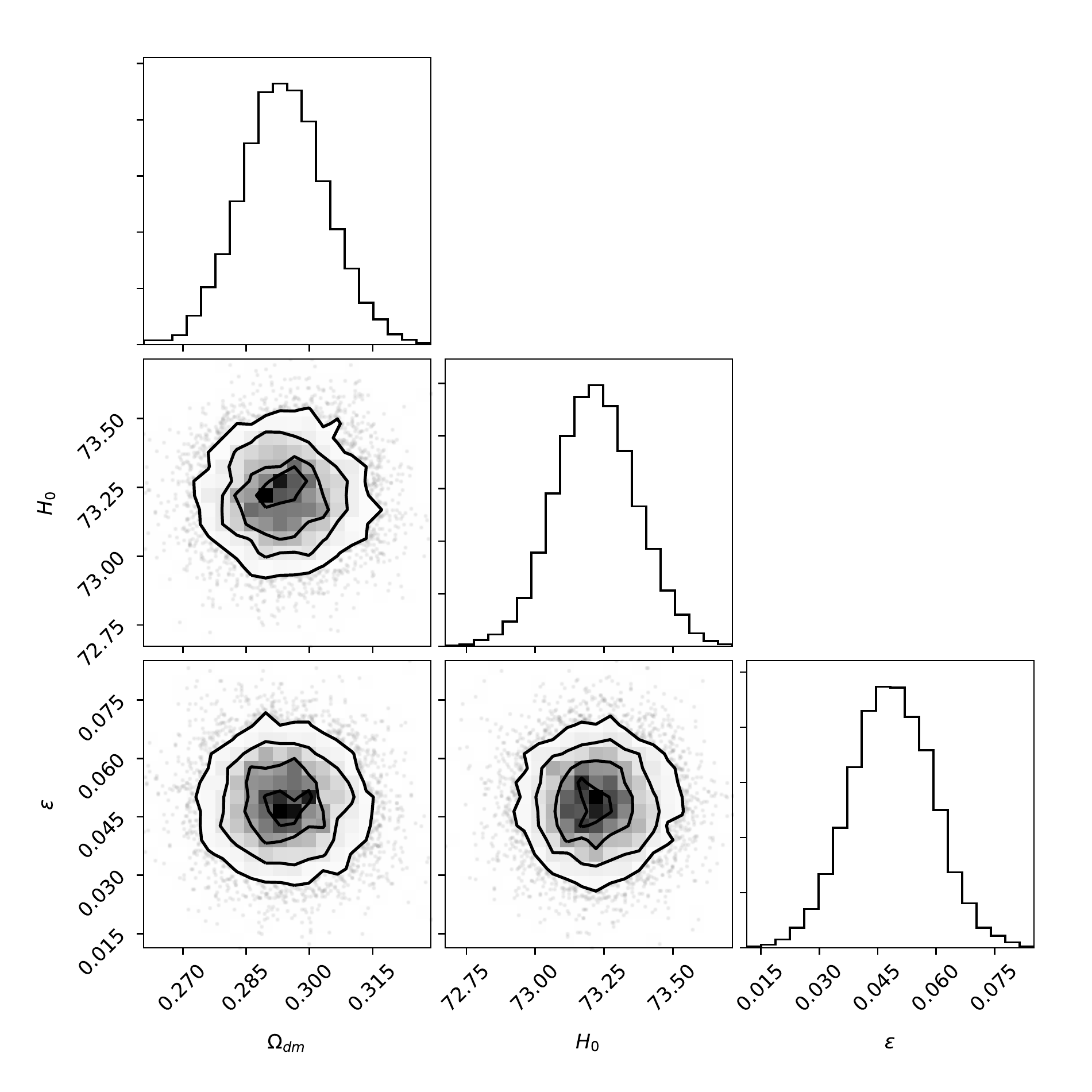}
\includegraphics[width=90 mm]{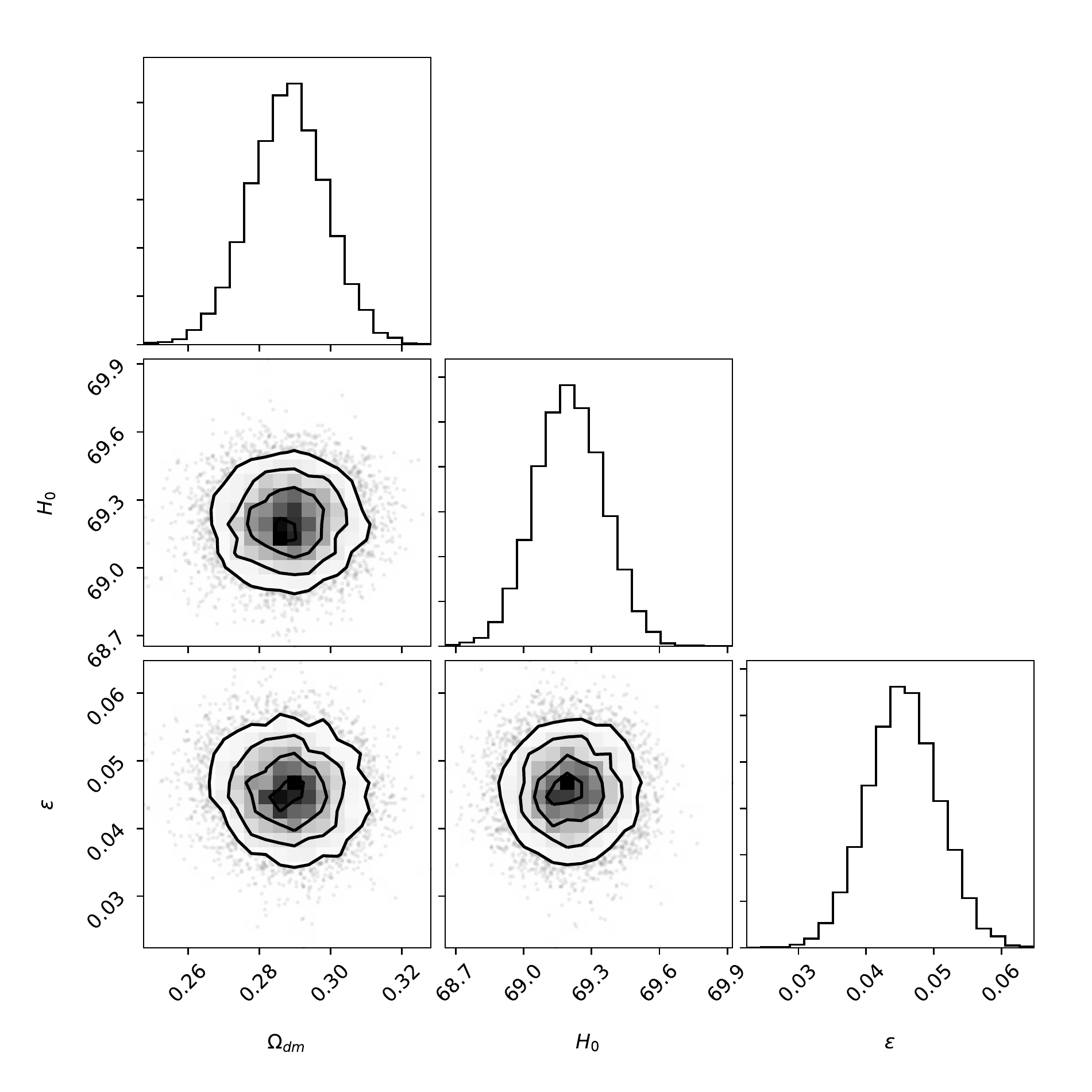}
 \end{array}$
 \end{center}
\caption{The contour map of the model given by Eq.~(\ref{eq:tau2}) for $z \in [0,2.5]$ is given by the left hand side plot, while the right hand side one depicts the contour map of the same model for $z \in [0,5]$. The best fit values of the model parameters, when $\tau(z)$ is given by  Eq.~(\ref{eq:tau2}), are $\Omega_{dm} = 0.294 \pm 0.098$, $H_{0} = 73.22 \pm 0.145$ and $\epsilon = 0.0486 \pm 0.0105$ for $z\in[0,2.5]$, while for $z \in [0,5]$ they are found to be $\Omega_{dm} = 0.288 \pm 0.011$, $H_{0} = 69.20 \pm 0.15$ and $\epsilon = 0.0455 \pm 0.0053$. The theoretical calculations of the luminosity distance are for the $\Lambda$CDM model.}
 \label{fig:Fig1_2}
\end{figure}

On the other hand, when we consider the second model for $\tau(z)$, Eq.~(\ref{eq:tau2}), we get the following results:

\begin{itemize}

\item  $\Omega_{dm} = 0.294 \pm 0.098$, $H_{0} = 73.22 \pm 0.145$ and $\epsilon = 0.0486 \pm 0.0105$ for $z\in[0,2.5]$, indicating again a non-transparent Universe. Also, we get higher values for the means of $\Omega_{dm}$ and $H_{0}$, as compared with the case for $\tau(z)$ being given by Eq.~(\ref{eq:tau1}). If in the previous case with $\tau(z)$ given by Eq.~(\ref{eq:tau1}, the $H_{0}$ tension problem could not be solved, here just the mean of $H_{0}$ is already enough to provide a hint to its solution.  The contour map corresponding to this case is shown in Fig.~(\ref{fig:Fig1_2})~(left-hand side plot).

\item An interesting result was obtained when we considered the  $z\in[0,5]$ redshift range. Indeed,  we wound that $\Omega_{dm} = 0.288 \pm 0.011$, $H_{0} = 69.20 \pm 0.15$ and $\epsilon = 0.0455 \pm 0.0053$. 
The contour map corresponding to this case is depicted in Fig.~(\ref{fig:Fig1_2})~(right-hand side plot).

\item On the other hand, the constraints $\Omega_{dm} = 0.296 \pm 0.01$, $H_{0} = 71.23 \pm 0.15$ and $\epsilon = 0.04 \pm 0.001$ obtained for $z\in[0,10]$ point again towards a non-transparent Universe as leading to a $H_{0}$ tension problem. The contour map corresponding to this case can be found in Fig.~(\ref{fig:Fig1_3})~(right-hand side plot).

\end{itemize}

To end this subsection, let us briefly recapitulate what we got. Using Bayesian (Probabilistic) Machine Learning based on ta data generative process with $D^{2}_{L,obs} = D^{2}_{L,true} e^{\tau}$, we have found that, provided $\Lambda$CDM is the correct cosmological model and the luminosity distance can be calculated through Eq.~(\ref{eq:DLTrue}), then it is most likely that we live in a non-transparent Universe. And the second very important message coming out of this analysis is that, depending on the redshift range and the form of $\tau(z)$, the $H_{0}$ tension may remain as a problem or will just disappear. Eventually, our study demonstrates also that if we deal with cosmic opacity in a non-proper way, then we can face  some misleading interpretations. In Table~\ref{tab:Table1}, for the benefit of the reader, we summarize all our results of this section.

\begin{table}
  \centering
    \begin{tabular}{ | c | c | c | c |  p{2cm} |}
    \hline
    
  $\tau(z) = 2 \epsilon z$ & $H_{0}$ & $\Omega_{dm}$ & $\epsilon$  \\
      \hline
      
 when $z\in[0,2.5]$ & $72.23 \pm 0.15$ km/s/Mpc & $0.276 \pm 0.01$ & $-0.04 \pm 0.005$\\
          \hline
          
when $z\in[0,5]$  & $73.23 \pm 0.146$ km/s/Mpc& $0.274 \pm 0.01$ & $0.0602 \pm 0.001$ \\
         \hline
         
when $z\in[0,10]$  & $73.26 \pm 0.144$ km/s/Mpc& $0.289 \pm 0.009$ & $0.099 \pm 0.001$\\
           \hline
 
 \multicolumn{3}{c}{} \\ \hline
 
$\tau(z) = (1+z)^{2\epsilon} -1$ & $H_{0}$ & $\Omega_{dm}$ & $\epsilon$ \\
       \hline

when $z\in[0,2.5]$ & $73.22 \pm 0.145$ km/s/Mpc & $0.294 \pm 0.098$ & $0.0486 \pm 0.0105$  \\
         \hline
 when $z\in[0,5]$  & $69.20 \pm 0.15$ km/s/Mpc & $0.288 \pm 0.011$ & $0.0455 \pm 0.0053$  \\   
 
          \hline
when $z\in[0,10]$  & $71.23 \pm 0.15$ km/s/Mpc& $0.296 \pm 0.01$ & $0.04 \pm 0.001$\\

     \hline 
    
    \end{tabular}
\caption{Best fit values and $1\sigma$ errors estimated for Model 1, Eq.~(\ref{eq:tau1}), and for Model 2, Eq.~(\ref{eq:tau2}), for $z \in [0,2.5]$, $z \in [0,5]$, and $z \in [0,10]$, respectively. The results are obtained for the $\Lambda$CDM model from our Bayesian Learning approach, where the generative based process  uses $D^{2}_{L,obs} = D^{2}_{L,true} e^{\tau}$, with $D_{L,true}$  given by Eq.~(\ref{eq:DLTrue}).}
  \label{tab:Table1}
\end{table}

\subsection{Cosmic opacity with XCDM}

\begin{table}
  \centering
    \begin{tabular}{ | c | c | c | c |  p{2cm} |}
    \hline
    
  $\tau(z) = 2 \epsilon z$ & $H_{0}$ & $\Omega_{dm}$ & $\epsilon$ & $\omega_{de}$ \\
      \hline
      
 when $z\in[0,2.5]$ & $68.28 \pm 0.15$ km/s/Mpc & $0.29 \pm 0.0097$ & $0.033 \pm 0.0051$ & $-0.96 \pm 0.01$\\
          \hline
          
when $z\in[0,5]$  & $70.22 \pm 0.15$ km/s/Mpc& $0.272 \pm 0.012$ & $0.0603 \pm 0.001$ & $-1.1 \pm 0.00098$\\
         \hline
         
when $z\in[0,10]$  & $69.26 \pm 0.149$ km/s/Mpc& $0.289 \pm 0.0095$ & $0.0994 \pm 0.0095$ & $-0.99 \pm 0.001$\\
           \hline
 
 \multicolumn{4}{c}{} \\ \hline
 
$\tau(z) = (1+z)^{2\epsilon} -1$ & $H_{0}$ & $\Omega_{dm}$ & $\epsilon$  & $\omega_{de}$\\
       \hline

when $z\in[0,2.5]$ & $68.21 \pm 0.151$ km/s/Mpc & $0.29 \pm 0.001$ & $0.04 \pm 0.001$  & $-0.962 \pm 0.009$\\
         \hline
 when $z\in[0,5]$  & $71.21 \pm 0.149$ km/s/Mpc & $0.29 \pm 0.01$ & $0.1 \pm 0.001$ & $ -1.13 \pm 0.05$ \\   
 
          \hline
when $z\in[0,10]$  & $71.23 \pm 0.15$ km/s/Mpc& $0.296 \pm 0.01$ & $0.1 \pm 0.001$ & $-1.11 \pm 0.048$\\

     \hline 
    
    \end{tabular}
\caption{Best fit values and $1\sigma$ errors estimated for Model 1, Eq.~(\ref{eq:tau1}), and for Model 2, Eq.~(\ref{eq:tau2}), for $z \in [0,2.5]$, $z \in [0,5]$, and $z \in [0,10]$, respectively. The results has been obtained for the XCDM model from our Bayesian Learning approach, where the generative based process  uses $D^{2}_{L,obs} = D^{2}_{L,true} e^{\tau}$, with $D_{L,true}$ given by Eq.~(\ref{eq:DLTrue}).}
  \label{tab:Table2}
\end{table} 

In the previous subsection we discussed about the constraints obtained for the $\Lambda$CDM standard model. In this one we will discuss the obtained results corresponding to the XCDM cosmological model where the cosmological constant has been replaced by a dark energy model with $\omega_{de} \neq -1$. In other words, in this case we need to fit $4$ parameters instead $3$, as in the previous one. We start our discussion in the case when $\tau(z) = 2\epsilon z$:

\begin{figure}[h!]
 \begin{center}$
 \begin{array}{cccc}
\includegraphics[width=90 mm]{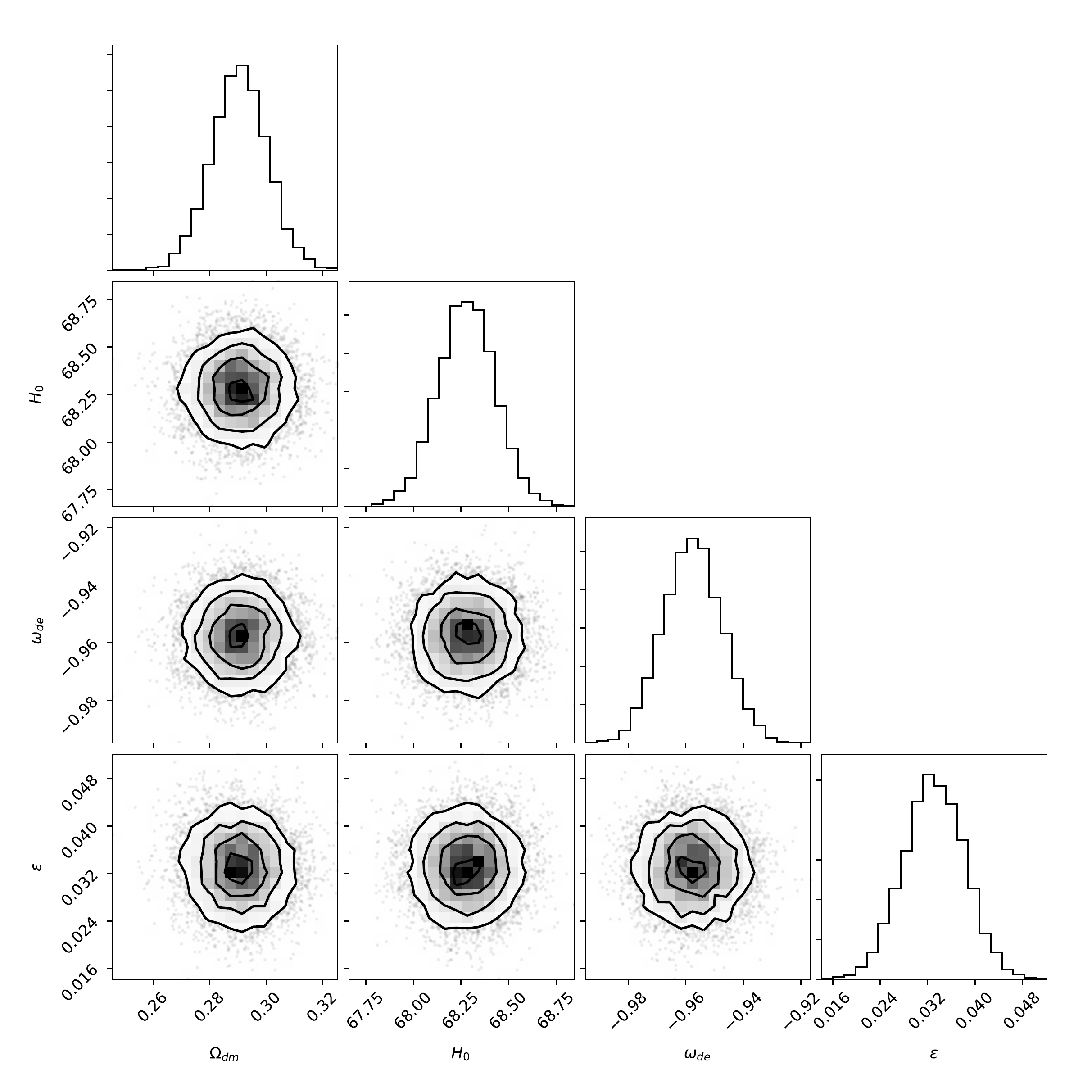}
\includegraphics[width=90 mm]{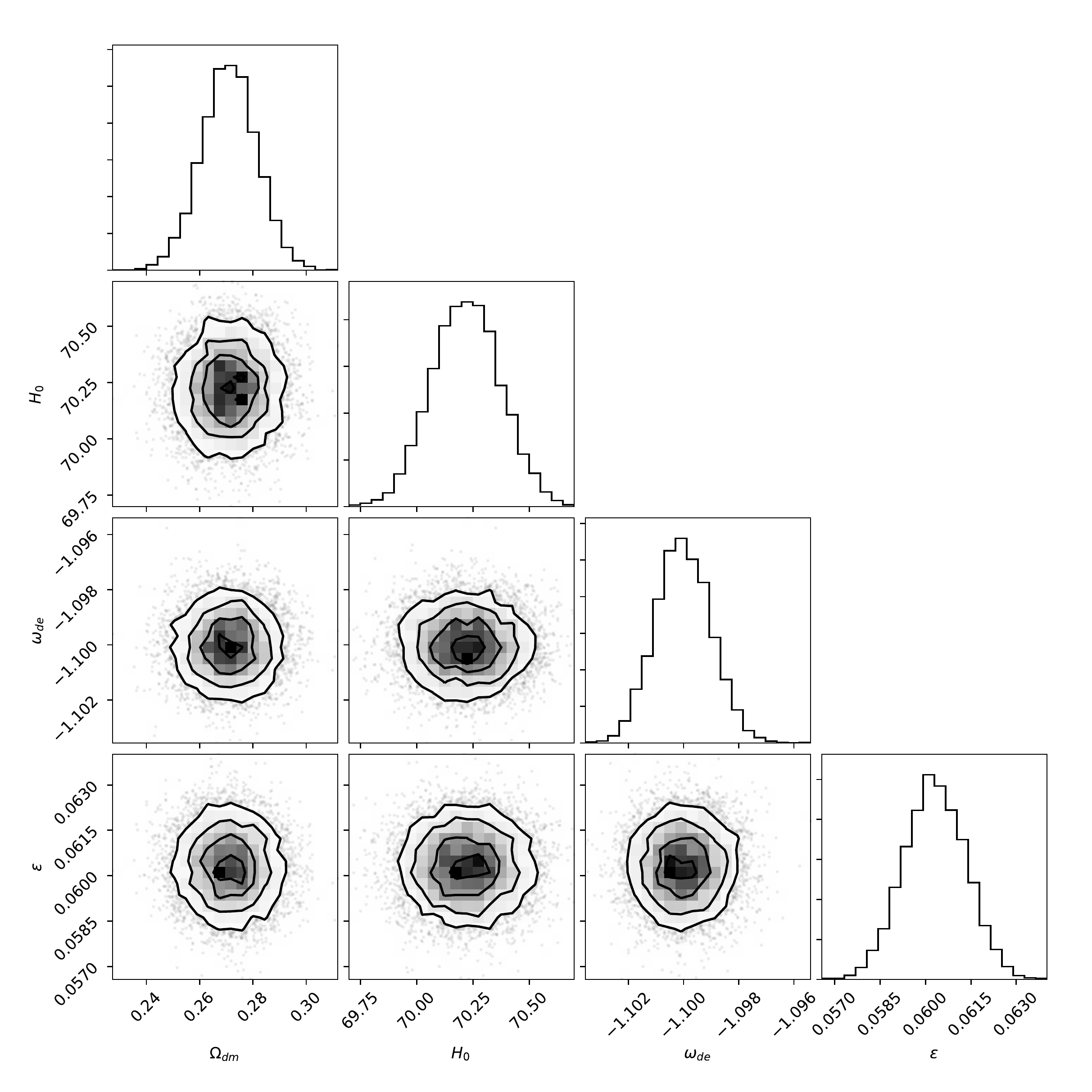}
 \end{array}$
 \end{center}
\caption{The contour map of the model given by Eq.~(\ref{eq:tau1}) for $z \in [0,2.5]$ is given by the left hand side plot, while the right hand side plot represents the contour map of the same model for $z \in [0,5]$. The best fit values of the model parameters, when $\tau(z)$ is given by  Eq.~(\ref{eq:tau1}), are $\Omega_{dm} = 0.29 \pm 0.0097$, $H_{0} = 68.28 \pm 0.15$, $\omega_{de} = -0.96 \pm 0.01$ and $\epsilon =  0.033 \pm 0.0051$ for $z\in[0,2.5]$, while for $z \in [0,5]$ they are found to be $\Omega_{dm} = 0.272 \pm 0.012$, $H_{0} = 70.22 \pm 0.15$, $\omega_{de} = -1.1 \pm 0.00098$ and $\epsilon = 0.0603 \pm 0.001$. The theoretical calculations of the luminosity distance are for the XCDM model.}
 \label{fig:Fig2_1}
\end{figure}

\begin{figure}[h!]
 \begin{center}$
 \begin{array}{cccc}
\includegraphics[width=90 mm]{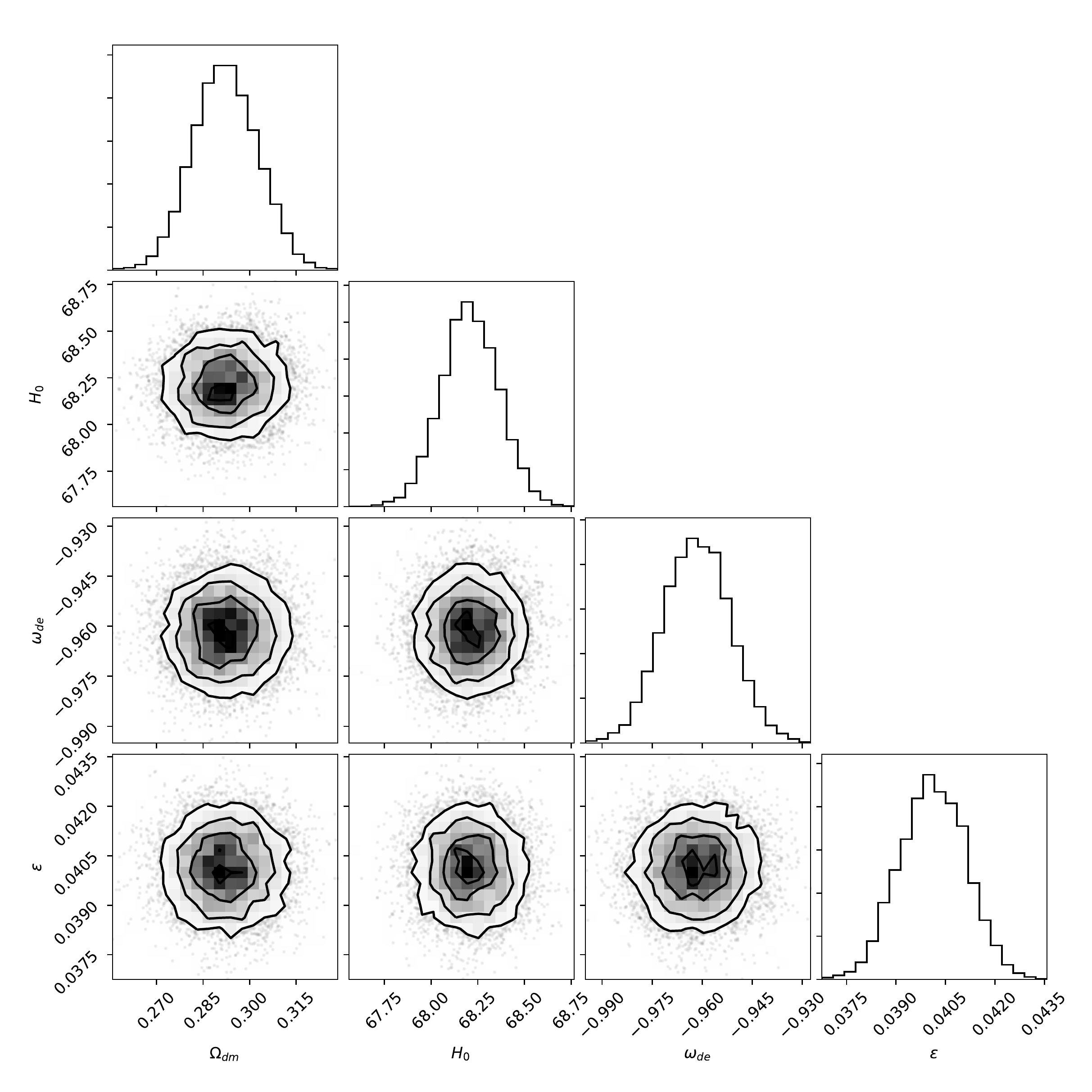}
\includegraphics[width=90 mm]{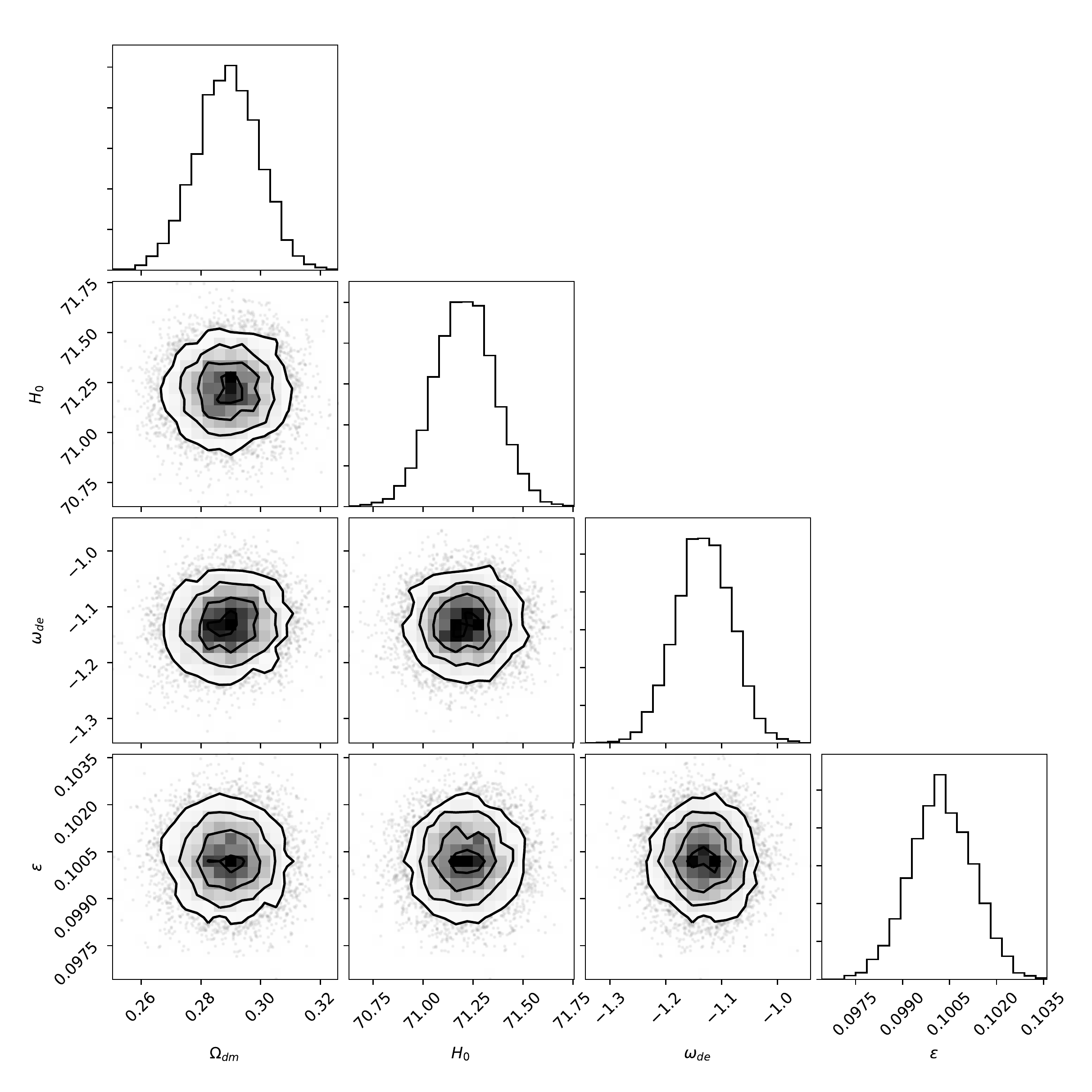}
 \end{array}$
 \end{center}
\caption{The contour map of the model given by Eq.~(\ref{eq:tau2}) for $z \in [0,2.5]$ is given by the left hand side plot, while the right hand side one shows the contour map of the same model for $z \in [0,5]$. The best fit values of the model parameters, when $\tau(z)$ is given by  Eq.~(\ref{eq:tau2}), are $\Omega_{dm} = 0.29 \pm 0.001$, $H_{0} = 68.21 \pm 0.151$, $\omega_{de} = -0.962 \pm 0.009$ and $\epsilon =  0.04 \pm 0.001$ for $z\in[0,2.5]$, while for $z \in [0,5]$ they are found to be $\Omega_{dm} = 0.29 \pm 0.01$, $H_{0} = 71.21 \pm 0.149$, $\omega_{de} = -1.13 \pm 0.05$ and $\epsilon = 0.1 \pm 0.001$. The theoretical calculations of the luminosity distance are for the XCDM model.}
 \label{fig:Fig2_2}
\end{figure}

\begin{itemize}

\item  In particular, we get $\Omega_{dm} = 0.29 \pm 0.0097$, $H_{0} = 68.28 \pm 0.15$, $\omega_{de} = -0.96 \pm 0.01$ and $\epsilon =  0.033 \pm 0.0051$ for $z\in[0,2.5]$.  The first noticeable result  is the  constrain for $H_{0}$. Indeed, as compared with the $\Lambda$CDM case, we have obtained a significantly reduced mean value for $H_{0}$. On the other hand, we geta a quintessence Universe that cannot be fully transparent.  The contour map corresponding to this case is given in Fig.~(\ref{fig:Fig2_1})~(left-hand side plot).

\item Moreover,  the Bayesian (Probabilistic) Machine Learning approach puts $\Omega_{dm} = 0.272 \pm 0.012$, $H_{0} = 70.22 \pm 0.15$, $\omega_{de} = -1.1 \pm 0.00098$ and $\epsilon = 0.0603 \pm 0.001$ constraints if $z\in[0,5]$. Our results indicate that the same XCDM model will led to a phantom Universe. And, again, it will not be transparent and we can have slightly higher value for the $H_{0}$ mean. In this analysis we saw also that the increase of the redshift range can increase $\epsilon$. Moreover, we have phantom dark energy. The contour map corresponding to this case can be found in Fig.~(\ref{fig:Fig2_1})~(the right-hand side plot).

\item On the other hand,  the obtained constraints $\Omega_{dm} = 0.289 \pm 0.0095$, $H_{0} = 69.26 \pm 0.149$, $\omega_{de} = -0.99 \pm 0.001$ and $\epsilon = 0.0994 \pm 0.0095$ when $z\in[0,10]$ shows that the Universe is not fully transparent and that we have a phantom dark energy. The contour map corresponding to this case can be found in Fig.~(\ref{fig:Fig2_3})~(left-hand side plot).

\end{itemize}

Before summarizing our observations for this part of the study we now briefly present the results when $\tau(z)$ is given by Eq.~(\ref{eq:tau2}). The constraints obtained are:

\begin{itemize}

\item $\Omega_{dm} = 0.29 \pm 0.001$, $H_{0} = 68.21 \pm 0.151$, $\omega_{de} = -0.962 \pm 0.009$ and $\epsilon =  0.04 \pm 0.001$ for $z\in[0,2.5]$. The contour map corresponding to this case is depicted in Fig.~(\ref{fig:Fig2_2})~(right-hand side plot).

\item $\Omega_{dm} = 0.29 \pm 0.01$, $H_{0} = 71.21 \pm 0.149$, $\omega_{de} = -1.13 \pm 0.05$ and $\epsilon = 0.1 \pm 0.001$. The contour map corresponding to this case is in Fig.~(\ref{fig:Fig2_2})~(right-hand side plot).

\item $\Omega_{dm} = 0.296 \pm 0.01$, $H_{0} = 71.23 \pm 0.15$, $\omega_{de} = -1.11 \pm 0.048$ and $\epsilon = 0.1 \pm 0.001$ constraints when $z\in[0,10]$. The contour map corresponding to this case is in Fig.~(\ref{fig:Fig2_3})~(right-hand side plot).

\end{itemize}

The results of our  study of the XCDM model for obtaining the luminosity distance associated to the generative process, using Bayesian (Probabilistic) Machine Learning have shown the following. First, that the Universe is actually not transparent and cannot at all be fully so. A second message is that the nature of dark energy can significantly influence the $H_{0}$ tension problem. Moreover, the opacity strongly affects our understanding of dark energy. This is not a completely new result; however, the way taken to confirm this fact is new. A question that arises from the comparison of the two results, for the $\Lambda$CDM and XCDM models, is the fact that they are model dependent, and future validation of the same can rely solely on actual observational data. 

However, Bayesian (Probabilistic) Machine Learning is also a probabilistic learning method and this means that during the generative process enough data were generated, allowing to obtain a feasibly good approach to the real world. Taking this into account, we can say that actually we explored the true nature of the models, and any deviation from presented results would indicate a tension in the observations. The last fact about the tension actually has been, and continues to be, neglected in standard Bayesian $\chi^{2}$ analysis, which indeed can lead to different challenges. Taking into account the observational limitations in the data, future discussion on this topic should be suspended, until these limitations can be possibly overcome.

In Table~\ref{tab:Table2} we summarize the results of our fit, which allows the reader to easily follow the above discussion of our analysis.

\section{Conclusions}\label{sec:conc}

Machine Learning algorithms have proven to be very efficient in analyzing large amounts of data in very different domains, and can significantly speed up the study of many different  processes in physics, chemistry, medicine, and biology. Behind Machine Learning algorithms, there are always three main steps, which can be formulated in the following way. One first  needs to define the model, then choose a set of data, and finally run a learning algorithm. Basically, the data are presented in terms of input output pairs, to train the neural network, and the data are coming from some experiment or observation. However, there is another interesting approach that can be used, instead of the usual Machine Learning algorithm,  known as Bayesian Machine Learning or Probabilistic Machine Learning. The steps to follow are again the same; however the definition of the data set has to be modified, it is conceptually different. In fact, real data are here replaced by the results coming out of some generative process,  directly connected to the model that we are going to study. In other words, Bayesian (Probabilistic) Machine Learning provides the appropriate tools allowing to study the model based on simulated/generated ``observational'' data. 

The dimming of type Ia supernovae~(SNe Ia), taken as standard candles in order to prove the accelerated expansion of the universe, may be actually affected by  at least four different sources of opacity. Indeed, it could be because of our galaxy's dust, that of the intergalactic medium, of the intervening galaxies, and of the host galaxy. It is a fact that, whether the Universe is transparent or not, will imprint a more or less significant effect on the parameter estimates of the model, and lead to misleading values, consequently affecting the result. Even if subsequent observations have independently confirmed the accelerating expansion of the Universe, we still need to estimate whether the Universe is transparent or not, for the fine details of the expansion parameter (the Hubble constant), in different domains of the universe, at different epochs of its evolution (redshifts ranges) will depend on this. 

In our study, we have applied Bayesian (Probabilistic) Machine Learning and probabilistic programming to constrain the cosmic opacity for two different models, namely, the flat $\Lambda$CDM one and a flat XCDM model. In the last one, dark energy is not given by the usual cosmological constant, but it is described by an $\omega_{de} \neq -1$ equation of state parameter.  In the generative process, we have used the fact that the flux that the observer gets, taking into account the cosmic opacity, will be reduced. The observed luminosity distance has the following form, $D^{2}_{L,obs} = D^{2}_{L,true} e^{\tau}$, where $\tau$ is the opacity parameter, while $D_{L,true}$ is the luminosity distance associated to the cosmological model. Constraints are obtained for the three redshift ranges: $z\in[0,2.5]$, $z \in[0,5]$, and $z\in[0,10]$, respectively. The last, wider redshift range has been considered in order to cover  recent and near future observations, too~(see, for instance, \cite{Inserra}~-~\cite{Sartoris}). On the other hand, the other two redshift ranges are necessary in order to understand how the constraints on the cosmic opacity can change, and also, if it is redshift-range dependent or not.  We adopted the following two forms, $\tau(z) = 2\epsilon z$ and $\tau(z) = (1+z)^{2\epsilon} -1$,  to denote the opacity as seen from an observer at $z=0$, coming from a source at $z$.

Namely, using Bayesian (Probabilistic) Machine Learning based on the data generation process with $D^{2}_{L,obs} = D^{2}_{L,true} e^{\tau(z)}$, we have found that if $\Lambda$CDM is the correct cosmological model, and the luminosity distance can be calculated by Eq.~(\ref{eq:DLTrue}), then our Universe is most likely non-transparent. And the second very important message coming out of this analysis is that, depending on the redshift range, the $H_{0}$ tension may still remain a problem or it can be explained and disappear. 

The other study using the XCDM model has led to the conclusion  that, in this case, the Universe is not transparent, and that it cannot be fully transparent for any range of redshifts. Another clear message from our study is that the nature of dark energy can significantly affect the $H_{0}$ tension problem. This is not a new result, however, the tools we have used to confirm this fact are new. This can be seen from the comparison of the results obtained for the two models considered, $\Lambda$CDM and XCDM. Moreover, we see that our understanding of the nature of dark energy can be significantly affected by the new constrains, indicating also a tendency to a clear departure from the standard cosmological constant case. On the other hand, depending on the model describing the opacity of the Universe, dark energy could have a  phantom nature, but even this fact could change depending on the redshift range studied. This could also be an indication that the specific form considered for $\tau(z)$ cannot correctly reflect the real nature of the cosmic opacity. On the other hand, another hint that follows from our study is that the CDD relation is not a proper way to study the cosmic opacity. The last, of course, requires future studies and the outcome will be reported in another paper.

An intriguing aspect behind our study is the fact that the results obtained are model dependent and  future validation of the same will rely solely on the new observational data to be obtained in future surveys. However, Bayesian (Probabilistic) Machine Learning is also a probabilistic learning method and this means that, during the generative process, enough data are generated, allowing to obtain a quite realistic behavior of the model. Taking all this into account, we are confident that we have actually explored the true nature of the models, and any deviation from the results here would most likely indicate a tension in the observations themselves. The last fact about the tension has actually been and continues to be neglected in standard Bayesian $\chi^{2}$ analysis, which indeed can cause different challenges. Taking into account the present observational limitations, future discussion on this issue should be postponed until these limitations can eventually be overcome. Finally, our strategy can be very useful for the construction of correct models for the cosmic opacity. This will be dealt with in a forthcoming project.

\section*{Acknowledgement}

This work has been partially supported by MINECO (Spain), project FIS2016-76363-P, and by AGAUR (Catalan Government), project 2017-SGR-247.

\section*{Appendix}

The nature of our study is quite different from that of other similar ones discussed in the recent literature. It has a predictive advantage, concerning the high redshift ranges, where data are not available. We think that, for the advantage of the readers, it is reasonable to report here all the details of the method. To validate the constraints obtained on the model parameters, we present all the contour maps of the models discussed in this work. 

\begin{figure}[h!]
 \begin{center}$
 \begin{array}{cccc}
\includegraphics[width=90 mm]{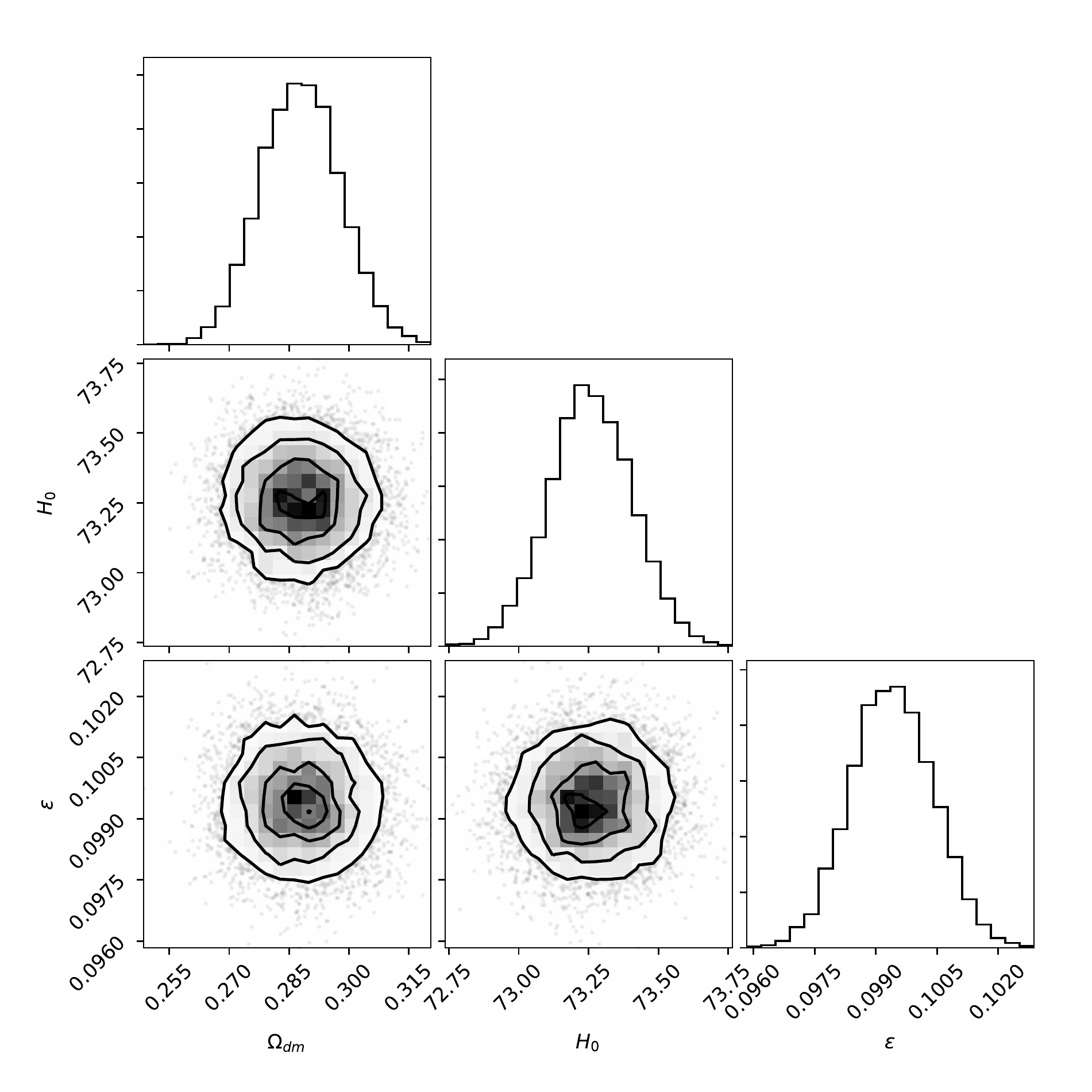}
\includegraphics[width=90 mm]{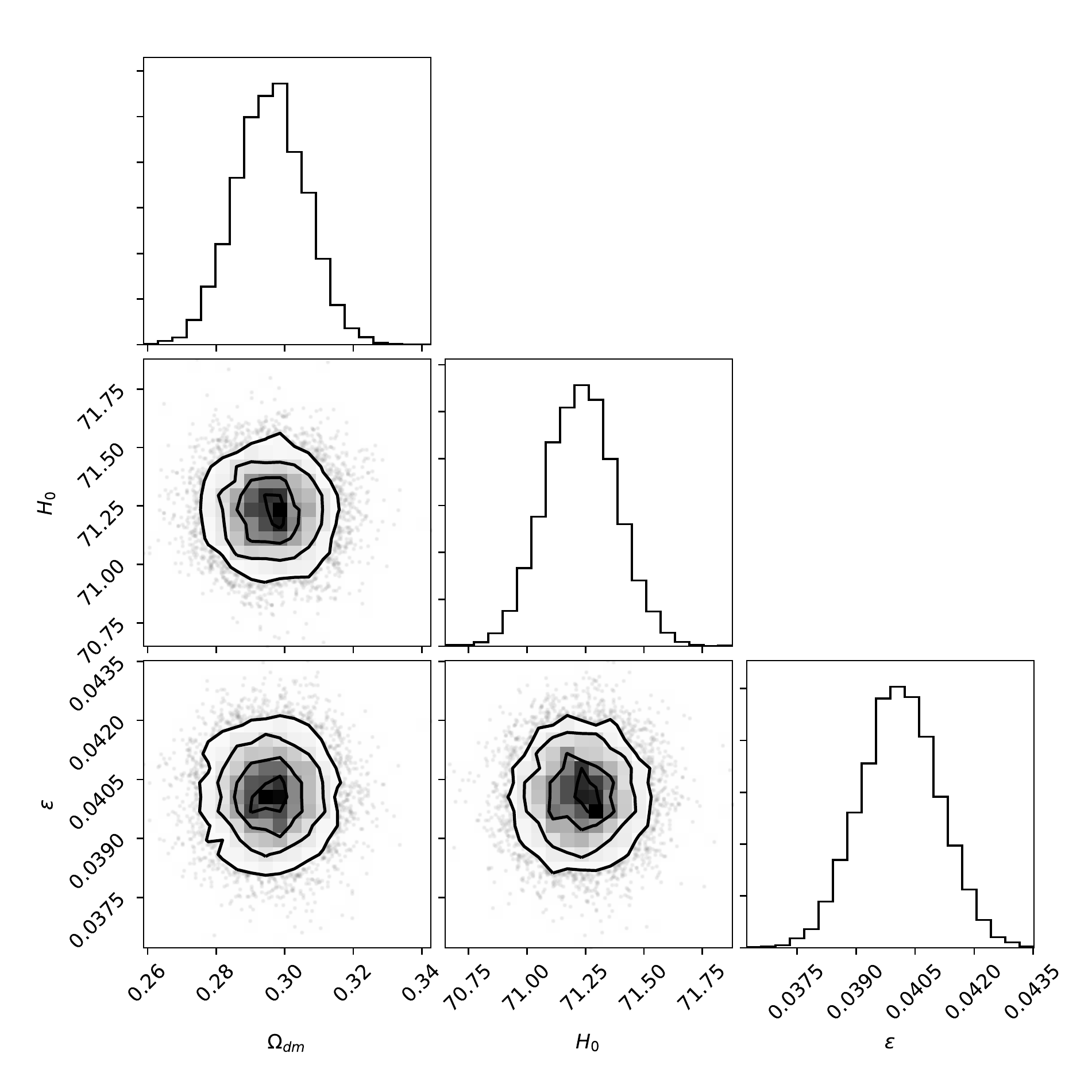}
 \end{array}$
 \end{center}
\caption{The contour map of the model given by Eq.~(\ref{eq:tau1}) for $z \in [0,10]$ is given by the left hand side plot, while the right hand side one represents the contour map of the model given by  Eq.~(\ref{eq:tau2}) for the same redshift range. The best fit values of the model parameters, when $\tau(z)$ is given by Eq.~(\ref{eq:tau1}), are $\Omega_{dm} = 0.289 \pm 0.009$, $H_{0} = 73.26 \pm 0.144$ and $\epsilon = 0.099 \pm 0.001$; while for the model given by Eq.~(\ref{eq:tau2}) the best fit values are $\Omega_{dm} = 0.296 \pm 0.01$, $H_{0} = 71.23 \pm 0.15$ and $\epsilon = 0.04 \pm 0.001$. The theoretical calculations of the luminosity distance are for the $\Lambda$CDM model.}
 \label{fig:Fig1_3}
\end{figure}

\begin{figure}[h!]
 \begin{center}$
 \begin{array}{cccc}
\includegraphics[width=90 mm]{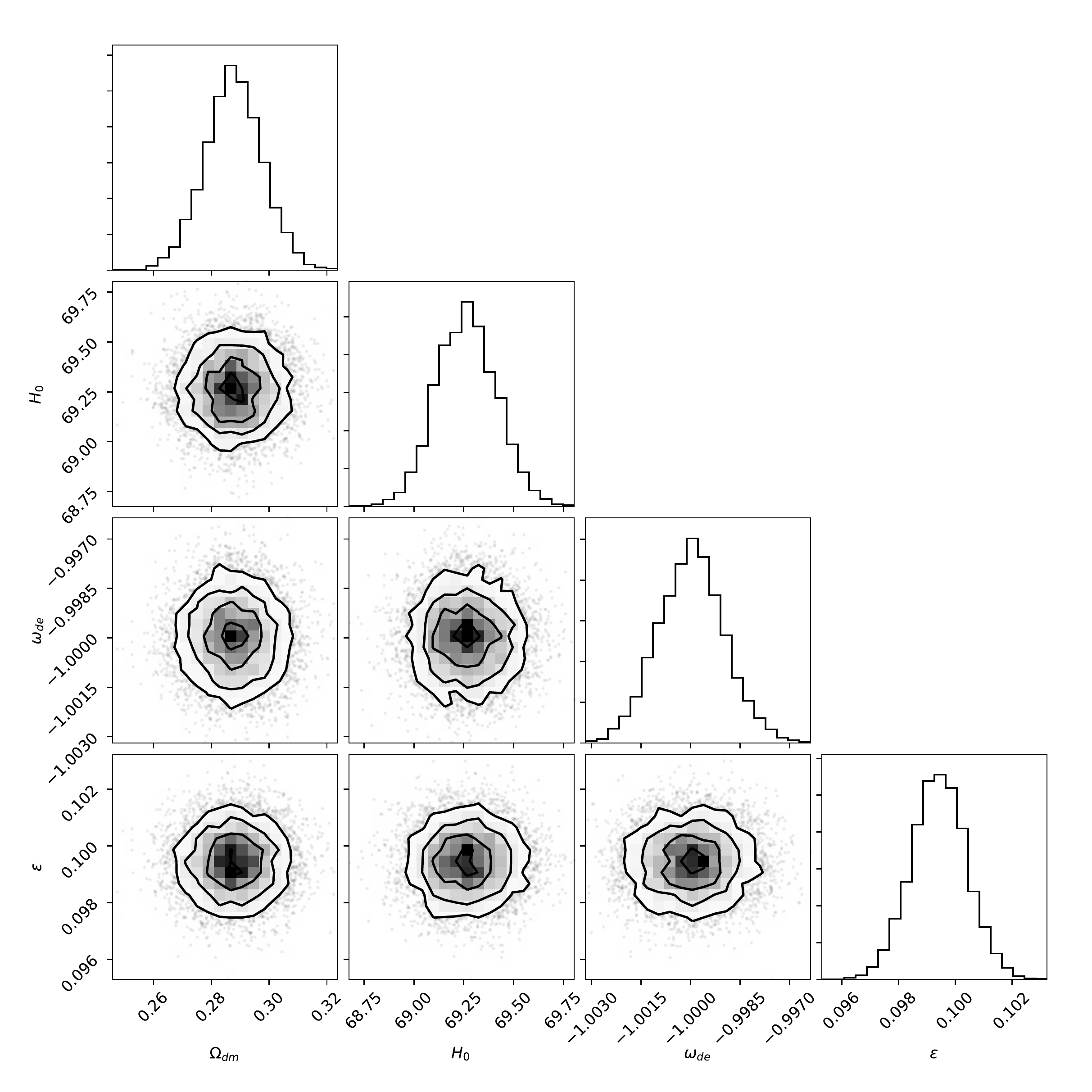}
\includegraphics[width=90 mm]{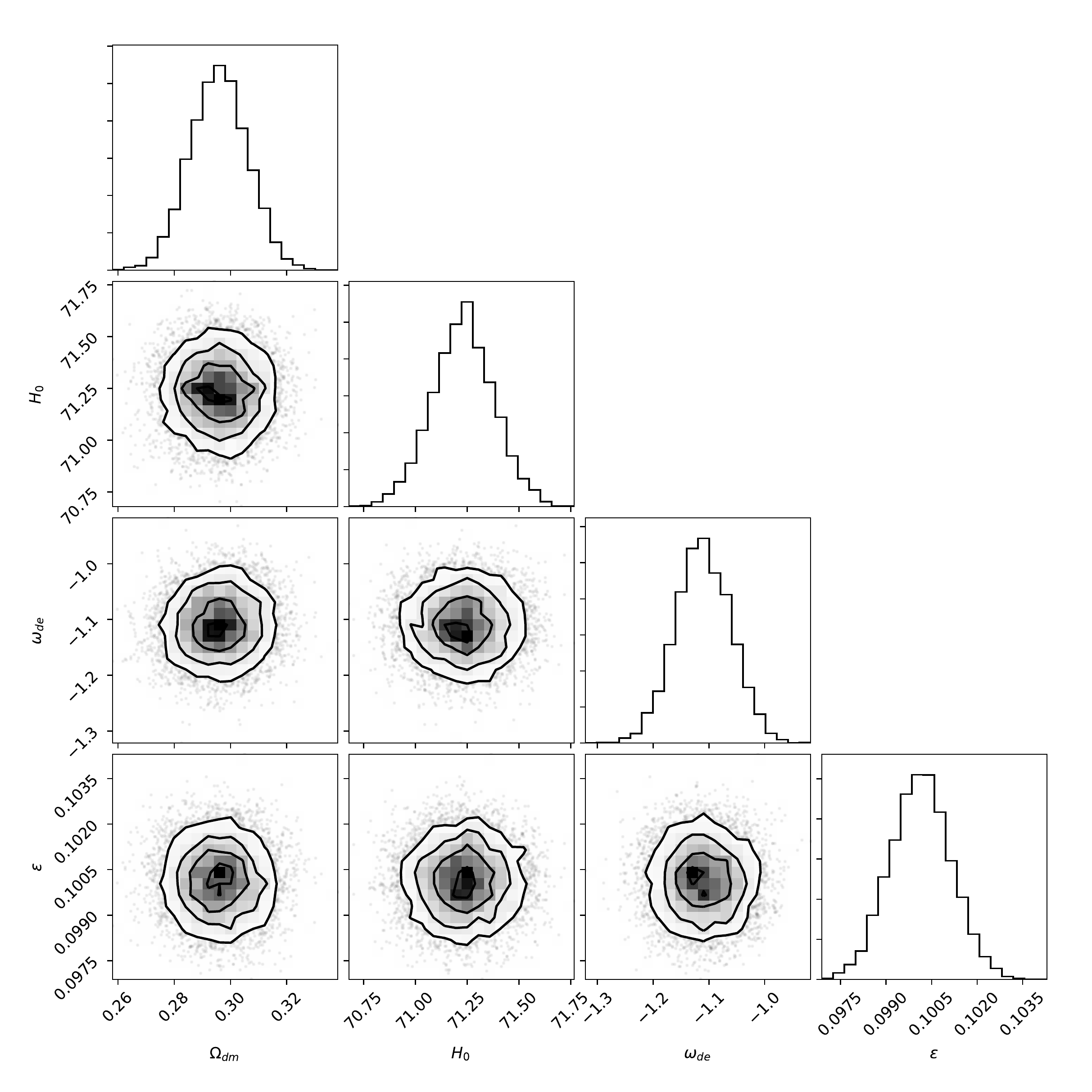}
 \end{array}$
 \end{center}
\caption{The contour map of the model given by Eq.~(\ref{eq:tau1}) for $z \in [0,10]$ is given by the left hand side plot, while the right hand side one corresponds to the model given by  Eq.~(\ref{eq:tau2}), for the same redshift range. The best fit values of the model parameters, when $\tau(z)$ is given by Eq.~(\ref{eq:tau1}), are $\Omega_{dm} = 0.289 \pm 0.0095$, $H_{0} = 69.26 \pm 0.149$, $\omega_{de} = -0.99 \pm 0.001$ and $\epsilon = 0.0994 \pm 0.0095$; while for the model given by Eq.~(\ref{eq:tau2}), the best fit values are $\Omega_{dm} = 0.296 \pm 0.01$, $H_{0} = 71.23 \pm 0.15$, $\omega_{de} = -1.11 \pm 0.048$ and $\epsilon = 0.1 \pm 0.001$. The theoretical calculations of the luminosity distance are for the XCDM model.}
 \label{fig:Fig2_3}
\end{figure}

In this Appendix we show the contour map of our models just for the extended redshift range, $z \in [0,10]$, in order to simplify the discussion. In particular, in Fig.~(\ref{fig:Fig1_3}) the contour map of the model given by Eqs.~(\ref{eq:tau1}) and  (\ref{eq:tau2}) is presented, where the theoretical calculations of the luminosity distance correspond to the $\Lambda$CDM model. On the other hand,  Fig.~(\ref{fig:Fig2_3}) corresponds to the contour map for the model given by Eqs.~(\ref{eq:tau1}) and (\ref{eq:tau2}), for the luminosity distance corresponding to the, more general, XCDM model. 

\end{document}